# Negative-U Centers as a Basis of Topological Edge Channels

Nikolay Bagraev[a], Eduard Danilovskii[a], Wolfgang Gehlhoff[b], Leonid Klyachkin[a], Andrey Kudryavtsev[a], Anna Malyarenko[a] and Vladimir Mashkov[c]

[a]*Ioffe Physical Technical Institute, Polytekhnicheskaya 26, 194021 St. Petersburg, Russia*
[b]*Technische Universitaet Berlin, D-10623, Berlin, Germany*
[c]*St. Petersburg State Polytechnical University, Polytekhnicheskaya 29, 195251 St. Petersburg, Russia*

**Abstract.** We present the findings of the studies of the silicon sandwich nanostructure that represents the high mobility ultra-narrow silicon quantum well of the *p*-type (Si-QW), 2 nm, confined by the δ-barriers, 3 nm, heavily doped with boron on the *n*-type Si (100) surface. The ESR studies show that nanostructured δ-barriers confining the Si-QW consist predominantly of the dipole negative-U centers of boron, which are caused by the reconstruction of the shallow boron acceptors along the <111> crystallographic axis, $2B^0 \rightarrow B^+ + B^-$. The electrically ordered chains of dipole negative-U centers of boron in the δ–barriers appear to give rise to the topological edge states separated vertically, because the value of the longitudinal, $G_{xx} = 4e^2/h$, and transversal, $G_{xy} = e^2/h$, conductance measured at extremely low drain-source current indicates the exhibition of the Quantum Spin Hall effect. Besides, the Aharonov-Casher conductance oscillations and the "$0.7 \cdot (2e^2/h)$-feature" obtained are evidence of the interplay of the spontaneous spin polarisation and the Rashba spin-orbit interaction that is attributable to the formation of the topological edge channels. We discuss the phenomenological model of the topological edge channel which can demonstrate the ballistic, Aharonov-Chasher effect or Josephson junction behaviour in dependence on the disorder in the distribution of the negative-U dipole centers in the upper and down δ–barriers.

**Keywords:** Quantum Spin Hall Effect, Double Barrier Structure, Edge Channels, Negative-U centers, Boron in Silicon.

**PACS:** 72.25.Dc, 72.20.-i

## INTRODUCTION

The investigation of edge spin-dependent transport is the subject of a considerable amount of research due to their applications in the modern physics direction – spintronics [1]. One of the best candidate on the role of such device that is able to demonstrate the topological helical edge channels with quantized conductance at high temperature appears to be the high mobility p-type silicon quantum well (Si-QW), 2 nm, confined by the δ-barriers heavily doped with boron [2]. Here the measurements of longitudinal and transversal voltage by top bias gating in the absence of the external magnetic field reveal the edge channels in such silicon nanosandwiches.

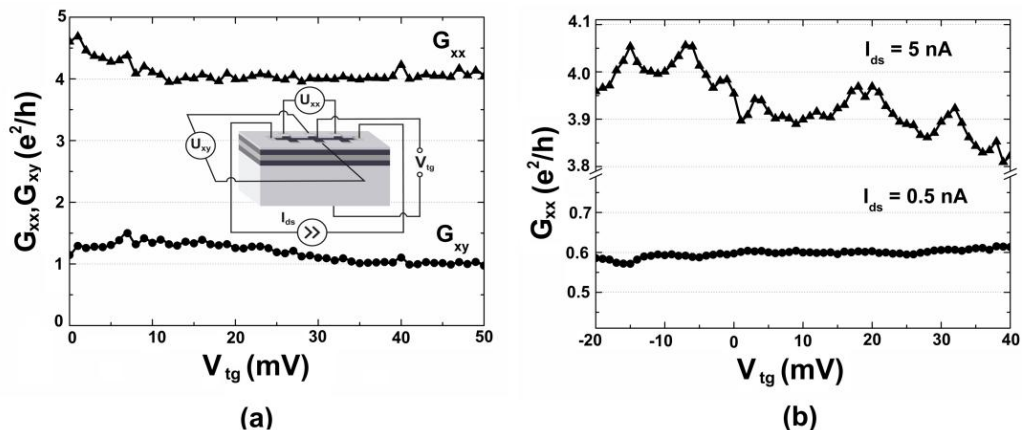

**FIGURE 1.** (a) The experimental dependence of the longitudinal, $G_{xx}$, and Hall, $G_{xy}$, conductance on the top gate voltage $V_{tg}$ at the stabilized drain-source current $I_{ds} = 0.25$ nA demonstrate the Quantum Spin Hall effect. T = 77 K. The inset in the middle of Fig.1 depicts the scheme of the experimental device. (b) Longitudinal conductance $G_{xx}$ as a function of the top gate voltage demonstrates the "$0.7 \cdot (2e^2/h)$-feature" at $I_{ds} = 0.5$ nA and the Aharonov–Casher oscillations at $I_{ds} = 5$ nA which are caused by the Rashba SOI changes.

# RESULTS AND DISCUSSION

The inset in the middle of Fig.1 represents the schematic diagram of the sandwich nanostructure device that demonstrates a perspective view of the p-type silicon quantum well (Si-QW), 2 nm, confined by the δ-barriers heavily doped with boron, $N_B = 5 \times 10^{21}$ cm$^{-3}$, on the n-type Si (100) surface. Thus, the δ - barriers represent really alternating arrays of the smallest undoped microdefects and doped dots with dimensions restricted to the value of 2 nm. The value of the boron concentration determined by the SIMS method seems to indicate that each doped dot located between undoped microdefects contains two impurity atoms of boron. Since the boron dopants form shallow acceptor centers in the silicon lattice, such high concentration has to cause a metallic-like conductivity. Nevertheless, in order to identify the edge states, the longitudinal and transversal (Hall) voltage were measured by varying the top gate bias voltage at different values of the highly-stabilized drain-source current in the silicon nanosandwiches. The longitudinal, $G_{xx} = 4e^2/h$, and transversal, $G_{xy} = e^2/h$, conductance, registered at extremely low value of the stabilized drain-source current, 0.25 nA, indicate the exhibition of the Quantum Spin Hall effect (Fig.1). Besides, the Aharonov-Casher conductance oscillations and the "$0.7 \cdot (2e^2/h)$-feature" obtained at small-scale drain-source current changes from 0.25 nA to 0.5 nA are evidence of the interplay of the spontaneous spin polarisation and the Rashba spin-orbit interaction that is attributable to the formation of the topological edge channels [2,3].

This conductivity properties of the δ - barriers between which the Si-QW is formed was quite surprising, when one takes into account the high level of their boron doping. To eliminate this contradiction, the ESR technique has been applied for the studies of the boron centers packed up in dots [4].

The angular dependences of the ESR spectra at different temperatures in the range 3.8÷27 K that reveal the trigonal symmetry of the boron dipole centers have been obtained with the Brucker-Physik AG ESR spectrometer at X-band (9.1-9.5 GHz) by the rotation of the magnetic field in the {110}-plane perpendicular to a {100}-interface ($B_{ext} = 0°$, 180° parallel to the Si-QW plane, $B_{ext} = 90°$ perpendicular to the Si-QW plane) (Fig. 2a, b, c and d). No ESR signals are observed, if the Si-QW confined by the δ - barriers is cooled down in the external magnetic field ($B_{ext}$) weaker than 0.22 T, with the persistence of the amplitude and the resonance field of the trigonal ESR spectrum as function of the crystallographic orientation and the magnetic field value during cooling down process at $B_{ext} \geq 0.22$ T (Fig. 2a, b and c). With increasing temperature, the ESR line observed changes its magnetic resonance field position and disappears at 27 K (Figure 2d).

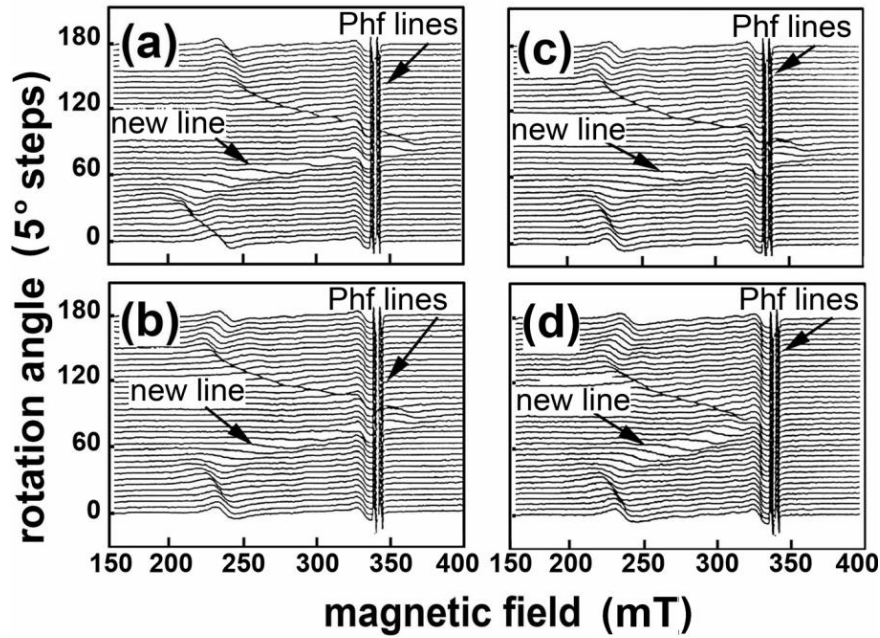

**FIGURE 2.** The trigonal ESR spectrum observed in field cooled ultra-shallow boron diffusion profile that seems to be evidence of the dynamic magnetic moment due to the trigonal dipole centers of boron inside the δ - barriers confining the Si-QW which is persisted by varying both the temperature and magnetic field values. $B_{ext} \parallel <110>$ (a), $\parallel <112>$ (b), $\parallel <111>$ (c, d). Rotation of the magnetic field in the {110}-plane perpendicular to a {100}-interface ($B_{ext} = 0°$, 180° $\parallel$ interface, $B_{ext} = 90°$ $\perp$ interface), ν = 9.45 GHz, T = 14 K (a, b, c) and T=21 K (d).

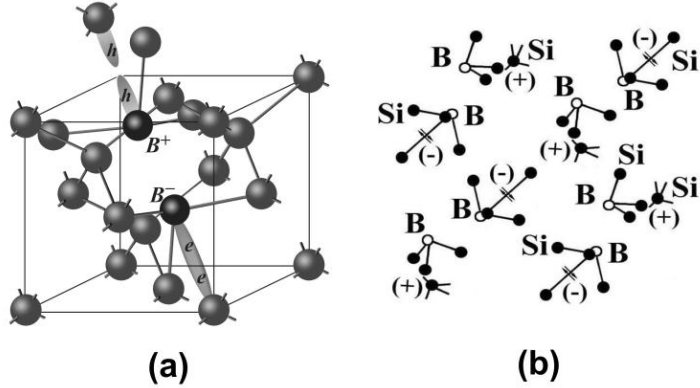

**FIGURE 3.** (a) Model for the elastic reconstruction of a shallow boron acceptor, which is accompanied by the formation of the trigonal dipole ($B^+$ - $B^-$) centers as a result of the negative-U reaction: $2B^o \rightarrow B^+ + B^-$. (b) A series of the dipole negative-U centers of boron located between the undoped microdefects that seem to be a basis of nanostructured δ - barriers confining the Si-QW.

The observation of the ESR spectrum is evidence of the fall in the electrical activity of shallow boron acceptors contrary to high level of boron doping. Therefore, the trigonal ESR spectrum observed seems to result from the dynamic magnetic moment that is induced by the exchange interaction between the small hole bipolarons which are formed by the negative-U reconstruction of the shallow boron acceptors, $2B^0 \rightarrow B^+ + B^-$, along the <111> crystallographic axis (Fig. 3a and b) [5]. These small hole bipolarons localized at the dipole boron centers, $B^+$ - $B^-$, seem to undergo the singlet-triplet transition in the process of the exchange interaction through the holes in the Si-QW thereby leading to the trigonal ESR spectrum (Fig. 2a, b, c and d).

Besides, electrostatically ordered dipole negative-U centers of boron seem to give rise to the topological superconductive edge states separated spatially in two δ–barriers. The electrical resistivity, magnetic susceptibility and specific heat measurements are actually evidence of the superconductor properties for the δ-barriers, $T_c$ =145 K, $2\Delta$ = 44 meV, $H_{c2}$ = 0.22 T [6]. The superconducting gap, $2\Delta$, that was also found by the local tunnelling spectroscopy (LTS) measurements has to reveal the THz emission due to the Josephson junctions self-assembled in the silicon sandwich nanostructure.

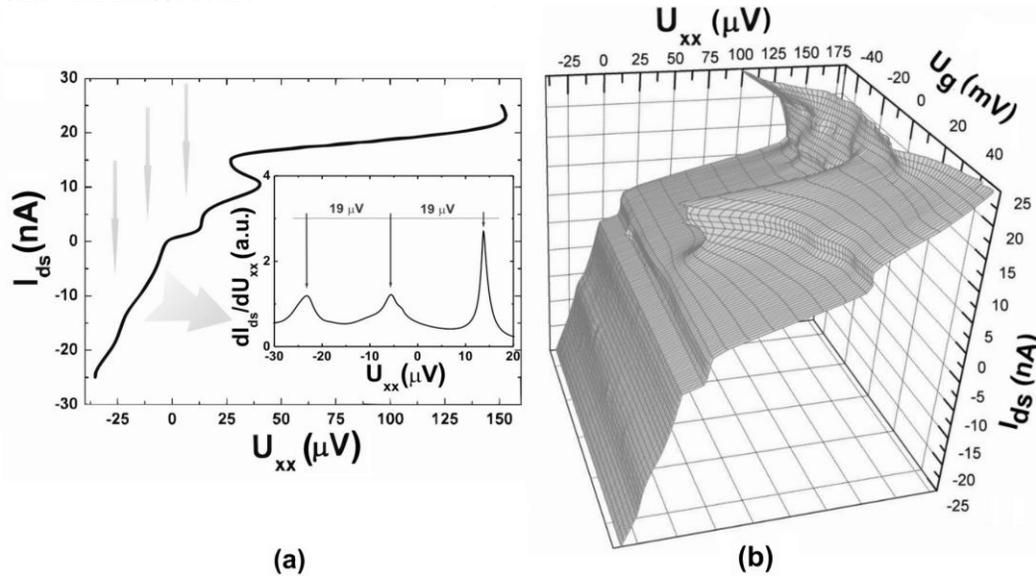

**FIGURE 4.** (a) The Fiske steps revealed by the $I_{ds}$-$U_{xx}$ dependence. Inset - the $dI/dU$-$U$ dependence. (b) The negative differential resistance controlled efficiently by varying the top gate voltage that verifies the GHz Josephson emission from the silicon sandwich nanostructure. $T$ = 77 K. The difference in the negative and positive $U_g$ effect is caused by the p-n junction presence in the device design. The absence of the CV characteristics symmetry appears to be due to disordering the dipole negative-U centers of boron which seem to give rise to the properties for topological insulator.

The HTS properties for the δ-barriers have been shown to result from the transfer of the small hole bipolarons through the negative-U dipole centers of boron, which cause the GHz generation under applied voltage or optical pumping [6]. This generation can be enhanced by introducing the internal microcavities in the Si-QW plane by varying the dimensions of the sandwich nanostructure using the photolithography technique. The dimensions of the sandwich nanostructure used in this work correspond to the formation of the 9.3 GHz microcavities.

The Fiske steps experiment was used to identify the Josephson generation revealed by measuring the CV characteristics of the device prepared (Fig. 4a and b). It is well known that the magnetic twist of the phase difference along the Josephson junction leads to the so-called Fiske steps [7]. These are nearly constant-voltage steps in the CV characteristic at voltages $V_n \approx (h/2e)c_s n/2L$, where $n = 1, 2,...$ and $L$ is the junction length perpendicular to the magnetic field, $c_s$ is the Swihart velocity of electromagnetic waves that are created by the Josephson generation and are able to propagate in the junction plane [8]. The Fiske steps in the CV characteristic and in the $dI/dU-U$ dependence are detected in the laboratory magnetic field by the potential $U_{xx}$ measurements under the conditions of the stepwise stabilization of the drain-source current, $I_{ds}$ (Figs. 4a and b). The Fiske steps observed are in agreement with the theory suggested by Kulik, with the Swihart velocity, $c_s$, equal to $1.6 \times 10^7$ m/s [8]. The negative differential resistance in the longitudinal CV characteristics, $I_{ds} > 0$, that results from the GHz Josephson emission appears to be controlled efficiently by varying the top gate bias voltage, $U_g$ (Fig. 4b). It necessary to be noticed, that the effect is maximal when the top gate voltage is equal to zero. This fact seems to be evidence of the weak disorder in the chain of the dipole boron centers by varying the top gate voltage that results in the asymmetry of the Josephson longitudinal CV characteristic in the presence of the p-n junction.

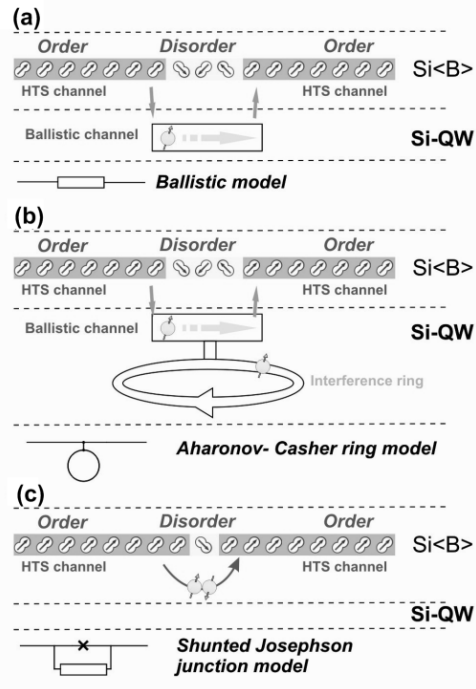

**FIGURE 5.** The models of the HTS topological channel which can demonstrate the ballistic (a), Aharonov-Chasher effect (b) or Josephson junction (c) behavior in dependence on the negative-U dipole centers disorder in the upper and down δ–barriers that is controlled by varying the longitudinal electric field value.

The high sensitivity of the longitudinal conductance to the drain-source current value appears to be due to disordering the dipole negative-U centers of boron by varying the external electrical field value which seem to give rise to the properties for the topological edge states separated spatially for the opposite spin orientation in the two δ-barriers. Topological insulators are electronic materials that have a bulk band gap like an ordinary insulator but have protected conducting states on their edge or surface. These states are theoretically possible due to the combination of spin-orbit interactions and time-reversal symmetry [1]. The topological edge channels appear to consist of the superconducting chains divided by weakly disordered dipole boron centers. Thus, the topological edge channels represent a series of parallel Josephson junctions shunted by ballistic conductors. Figs. 5 demonstrate examples of possible topological channel configurations. In dependence on external electrical field and edge channel quality it is appear to be realized the ballistic Si-QW topological channel or the ring shunted quantum dot contact resulting in the Aharonov-Casher effect registration or the ballistic shunted Josephson junction, (Figs. 5a, b and c, respectively). However, upper and down δ–barriers confining the Si-QW are differently disordered that is due to the device design, which results in the anisotropy of the spin-dependent transport of 1D holes.

## ACKNOWLEDGMENTS

The work was supported by the programme of fundamental studies of the Presidium of the Russian Academy of Sciences "Quantum Physics of Condensed Matter" (grant 9.12); programme of the Swiss National Science Foundation (grant IZ73Z0_127945/1); the Federal Targeted Programme on Research and Development in Priority Areas for the Russian Science and Technology Complex in 2007–2012 (contract no. 02.514.11.4074), the SEVENTH FRAMEWORK PROGRAMME Marie Curie Actions PIRSES-GA-2009-246784 project SPINMET.